\documentclass[%
 reprint,
 superscriptaddress,
 nobibnotes,
 amsmath,amssymb,
 aps,
 pra,
]{revtex4-2}

\usepackage{graphicx}
\usepackage{color}
\usepackage{dcolumn}
\usepackage{bm}
\usepackage{braket}
\usepackage{physics}
\usepackage{amsthm}
\usepackage{amssymb}
\usepackage{natbib}
\usepackage{siunitx}
\usepackage{booktabs}
\usepackage{mhchem}
\usepackage{threeparttable}
\usepackage{subfigure}
\usepackage{comment}
\usepackage[normalem]{ulem}

\usepackage{algorithm}
\usepackage{algpseudocode}

\theoremstyle{definition}

\theoremstyle{plain}

\theoremstyle{definition}

\theoremstyle{remark}

\begin{document}

%\preprint{APS/123-QED}

\title{Error suppression by a virtual two-qubit gate}

\author{Takahiro Yamamoto}%
\email{Takahiro.Yamamoto3@ibm.com}
\affiliation{%
 IBM Quantum, 19-21 Nihonbashi Hakozaki-cho, Chuo-ku, Tokyo, 103-8510, Japan
}%
\author{Ryutaro Ohira}%
\email{Ryutaro.Ohira@ibm.com}
\affiliation{%
 IBM Quantum, 19-21 Nihonbashi Hakozaki-cho, Chuo-ku, Tokyo, 103-8510, Japan
}%
\date{\today}

\begin{abstract}

Sparse connectivity of a superconducting quantum computer results in the large experimental overheads of SWAP gates. In this study, we consider employing a virtual two-qubit gate (VTQG) as an error suppression technique. The VTQG enables a non-local operation between a pair of distant qubits using only single qubit gates and projective measurements. Here, we apply the VTQG to the digital quantum simulation of the transverse-field Ising model on an IBM quantum computer to suppress the errors due to the noisy two-qubit operations. We present an effective use of VTQG, where the reduction of multiple SWAP gates results in increasing the fidelity of the output states. The obtained results indicate that the VTQG can be useful for suppressing the errors due to the additional SWAP gates. Additionally, by combining a pulse-efficient transpilation method with the VTQG, further suppression of the errors is observed. In our experiments, we have observed one order of magnitude improvement in accuracy for the quantum simulation of the transverse-field Ising model with 8 qubits.
\end{abstract}


\maketitle

%\tableofcontents

\section{Introduction}\label{Sec:introduction}

Qubit connectivity has a large impact on the performance of near term quantum computers. 
For instance, on a superconducting quantum device, an application of a two-qubit gate is limited on nearest-neighboring qubits, necessitating the SWAP gates for interactions between distant qubits. The SWAP gate is constructed by the concatenation of three CNOT gates. However, two-qubit gate operations are typically more prone to noise than single-qubit operations, thus, the requirement of the additional SWAP gates leads to significant reduction in the performance of the quantum device.

In this study, to suppress the errors due to the additional SWAP gates for a two-qubit gate operation between a pair of distant qubits, we have employed a virtual two-qubit gate (VTQG) proposed by Mitarai and Fuji \cite{mitarai2021constructing}. The original motivation behind the VTQG is to simulate a large quantum circuit with a small-scale quantum computer \cite{mitarai2021constructing}, and in this context, similar studies have been performed \cite{PhysRevLett.125.150504, PhysRevResearch.1.013006, PhysRevX.6.021043, tang2021cutqc, ying2022experimental}. 

Here, we have focused on the property that the VTQG allows a two-qubit gate between an arbitrary pair of qubits. This non-local operation reduces the number of additional SWAP gates since the communication between distant qubits otherwise requires a number of SWAP operations. In this sense, we believe that the VTQG can be used as an error suppression technique. In this study, we have demonstrated performance improvement of the quantum simulation of the transverse-field Ising model by implementing the VTQG on an IBM quantum device. The obtained results show that the VTQG can be a solid tool for suppressing errors. 

Furthermore, we have combined the VTQG with a pulse-efficient transpilation method \cite{PhysRevResearch.3.033171, PhysRevResearch.3.043088, ferris2022quantum, kim2021scalable, niu2022effects, melo2022pulse}. This pulse-efficient transpilation allows the reduction of the gate execution time as well as the number of the two-qubit gate operations. As demonstrated in the previous works \cite{PhysRevResearch.3.033171, PhysRevResearch.3.043088, ferris2022quantum, kim2021scalable, niu2022effects, melo2022pulse}, an enhancement of the performance of the quantum device can be expected.

\section{Methods}

In this section, we describe our model and the methodology used to perform the experiments on an IBM quantum device and the error suppression techniques used in the experiments. 

\subsection{Transverse-field Ising model}

\begin{figure}[t]
    \includegraphics[width=8.5cm]{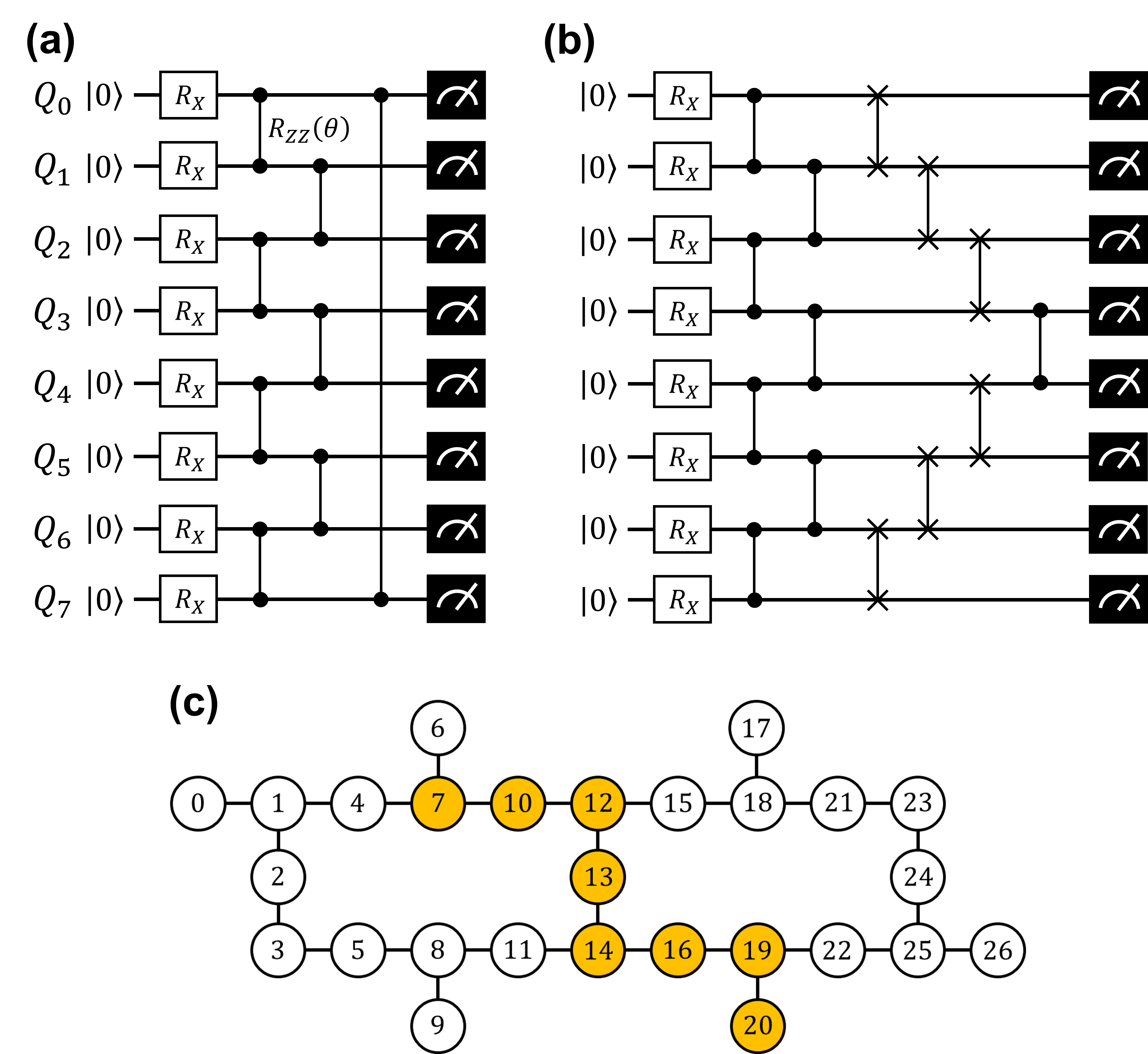}
    \caption{
    (a) An example of a quantum circuit with a single Trotter layer with 8 qubits to simulate the dynamics of the transverse Ising model. (b) Additional SWAP gates are required for applying an $R_{ZZ}(\theta)$ gate between the distant qubits. (c) A qubit layout of $ibmq$\rule{0.2cm}{0.15mm}$mumbai$. The qubits used in the experiments are highlighted in orange.
    \label{Fig:q_circuit} 
    }
\end{figure}

As a target problem, we study the dynamics of the one-dimensional transverse-field Ising model with the periodic boundary condition. The Hamiltonian of this model is given as follows:
\begin{equation}
    H = h\sum_i X_i - J\sum_i Z_iZ_{i+1} = H_X + H_{ZZ},
\end{equation}
where $h$ and $J$ correspond to the transverse magnetic field and the coupling strength between neighboring spins, respectively. The time evolution operator  of this model can be approximated by the first-order Trotter decomposition. Therefore, the dynamics of this model can be simulated on a quantum computer by applying the following unitary operators 
\begin{equation}
    e^{-iHt} \approx \prod^{n} e^{-iH_{ZZ} dt}e^{-iH_X dt},
\end{equation}
where $dt = t/n$ is time step and $n$ corresponds to the Trotter step. In Fig.\ref{Fig:q_circuit}(a), we show a typical quantum circuit which corresponds to a single Trotter layer with 8 qubits. The circuit to simulate the behaviour of the transverse-field Ising model consists of single-qubit rotation gates and two-qubit $R_{ZZ}(\theta)$ gates.

\subsection{Quantum hardware}

All experiments presented in this manuscript have been performed on a 27-qubit device $ibmq$\rule{0.2cm}{0.15mm}$mumbai$. The qubit layout of $ibmq$\rule{0.2cm}{0.15mm}$mumbai$ is given in Fig.\,\ref{Fig:q_circuit}(c). We run the circuits with up to 8 qubits, which are highlited in orange in Fig.\,\ref{Fig:q_circuit}(c). The average errors of a single-qubit gate and a CNOT gate applied to the 8 qubits are 0.03\% and 0.87\%, respectively. We kept records of the device property provided by IBM and took the average of them.

\subsection{Virtual two-qubit gate}

Depending on the qubit connectivity, the additional SWAP gates are required to apply a two-qubit operation between a pair of distant qubits. For instance, when we run the quantum circuit given in Fig.\,\ref{Fig:q_circuit}(a) using a qubit layout shown in Fig.\,\ref{Fig:q_circuit}(c), we apply 6 additional SWAP gates to apply an $R_{ZZ}$ to the first qubit ($Q_0$) and the $8$th qubit ($Q_{7}$) as shown in Fig.\,\ref{Fig:q_circuit}(b). However, the SWAP gate requires three CNOT gates, which are typically noisier than single-qubit operations. Therefore, as the problem size increases, the performance of the quantum device is significantly reduced due to the noisy SWAP gates. 

\begin{figure}[t]
    \includegraphics[width=8.5cm]{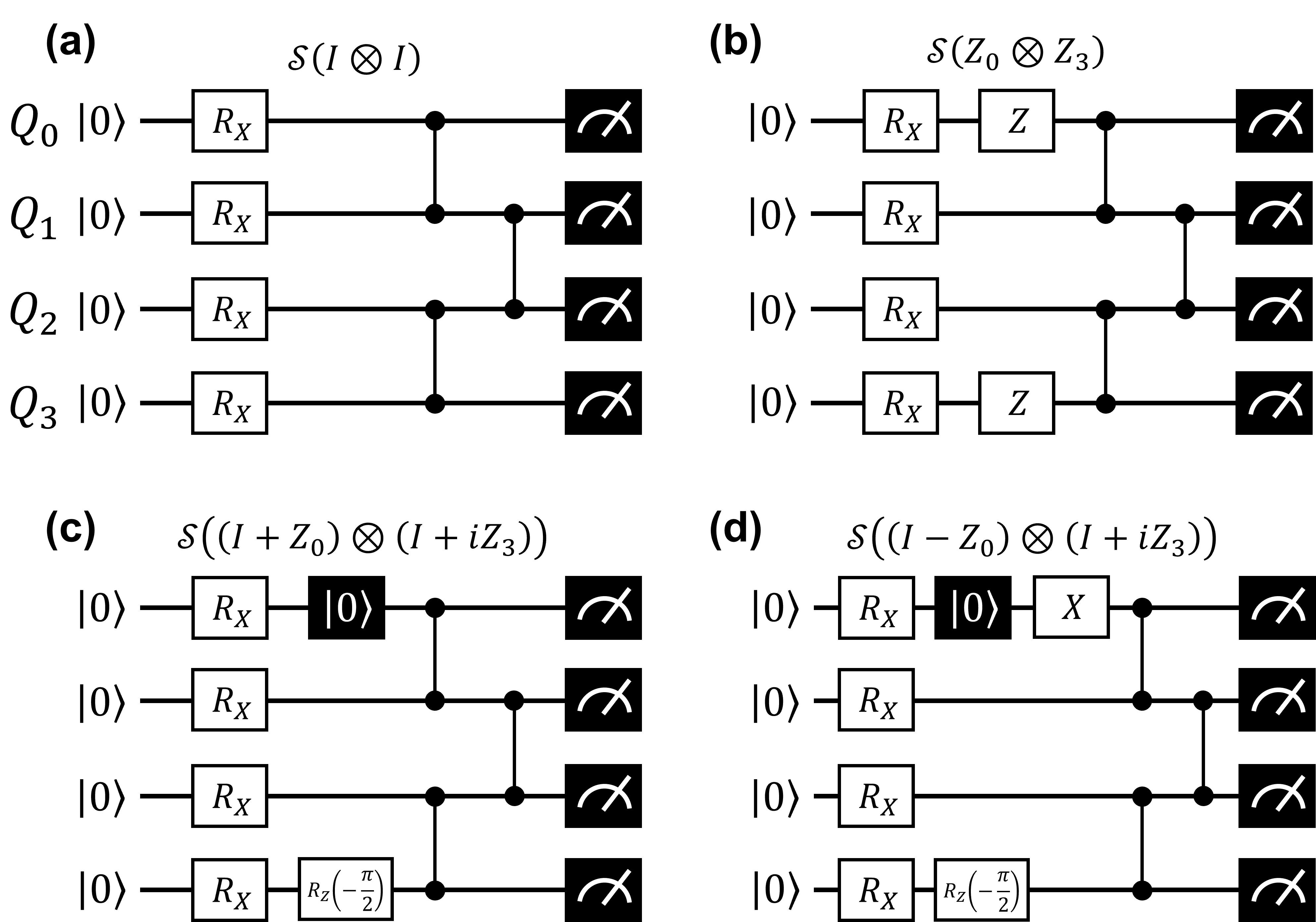}
    \caption{
    Quantum circuits for simulating the dynamics of the Ising model Hamiltonian using 4 qubits, where a VTQG is applied between $Q_0$ and $Q_{3}$. Each circuit implements (a) $\mathcal{S}(I_0 \otimes I_{3})$, (b) $\mathcal{S}(Z_0 \otimes Z_{3})$, (c) $\mathcal{S}((I + Z_0) \otimes (I + i Z_{3}))$, and (d) $\mathcal{S}((I - Z_0) \otimes (I + i Z_{3}))$, respectively.
    \label{Fig:hamiltonian_simulation_with_vtqg} 
    }
\end{figure}

We have replaced the $R_{ZZ}(\theta)$ applied between the first qubit ($Q_0$) and the $N$th qubit ($Q_{N-1}$) with a VTQG, which can be decomposed as
\cite{mitarai2021constructing}:
\begin{widetext}
\begin{equation}
\label{eq:vtqg-decomposition}
\begin{split}
&\mathcal{S}(e^{i \frac{\theta}{2} Z_0 \otimes Z_{N-1}}) = 
 \cos^2\frac{\theta}{2}\mathcal{S}(I \otimes I) 
+ \sin^2\frac{\theta}{2}\mathcal{S}(Z_0\otimes Z_{N-1}) \\ 
& + \frac{1}{8} \cos\frac{\theta}{2}\sin\frac{\theta}{2}
\sum_{\alpha_0, \alpha_{N-1}\in\qty{1,-1}} \alpha_0 \alpha_{N-1}
[\mathcal{S}((I+\alpha_0Z_0)\otimes(I+i\alpha_{N-1}Z_{N-1})) + \mathcal{S}((I+i\alpha_0Z_0)\otimes(I+\alpha_{N-1}Z_{N-1}))].
\end{split}
\end{equation}
\end{widetext}
Here $\mathcal{S}$ is a superoperator, where a unitary operation $U$ onto a state represented by density matrix $\rho$ is denoted as $\mathcal{S}(U) \rho = U\rho U^\dagger$. The operations $\mathcal{S}(I+\alpha Z)$ and $\mathcal{S}(I+i\alpha Z)$ for $\alpha \in\qty{1,-1}$ can be implemented by projective measurement on the $Z$ basis and single-qubit rotation about the $Z$ axis, respectively.
We show a set of the quantum circuits for simulating the dynamics of the Ising model Hamiltonian using a VTQG applied between $Q_0$ and $Q_{3}$ in Fig.~\ref{Fig:hamiltonian_simulation_with_vtqg}. 
The $R_{ZZ}(\theta)$ gate is implemented using the reset operations and the single qubit operations.
Other terms in Eq.\,(\ref{eq:vtqg-decomposition}) can be implemented in similar fashions, either by flipping the sign of the single-qubit rotation about the $Z$ axis or exchanging $Q_0$ and $Q_{3}$.

\begin{figure}[t]
    \includegraphics[width=8.5cm]{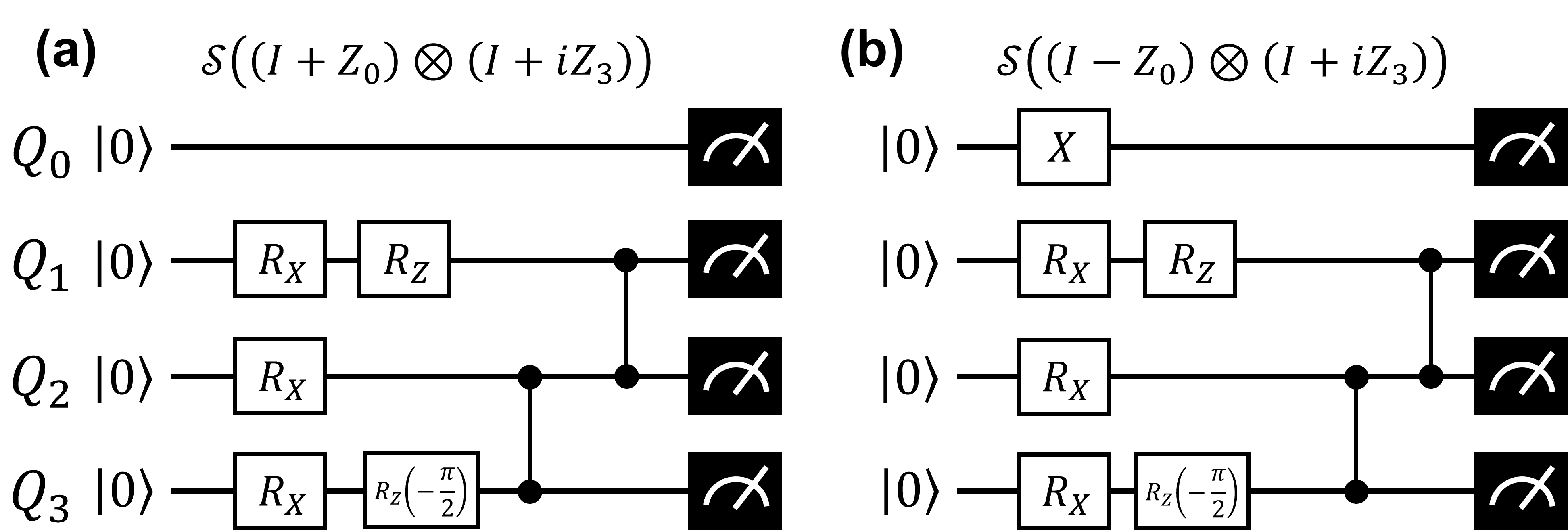}
    \caption{
    Simplified quantum circuits for implementing (a) $\mathcal{S}((I + Z_0) \otimes (I + i Z_{3}))$ and (b) $\mathcal{S}((I - Z_0) \otimes (I + i Z_{3}))$, respectively.
    \label{Fig:hamiltonian_simulation_with_vtqg_simplified} 
    }
\end{figure}

The quantum circuits for implementing $\mathcal{S}(I \pm Z)$ can be further simplified. We can remove the single-qubit rotation about the X axis proceeding the reset operation and classically calculate the probability 
\begin{equation}
    \mathrm{Tr} \left( \rho \frac{I + Z}{2} \right) = \cos^2 (\beta),
\end{equation}
where $\beta$ is the rotation angle. 
If the controlled qubit is in either $\ket{0}$ or $\ket{1}$, an $R_{ZZ}(\theta)$ gate simply works as an $R_Z(\theta)$ gate on the target qubit. 
Therefore, further simplification of the circuit shown in Figs.~\ref{Fig:hamiltonian_simulation_with_vtqg} (c) and (d) can be realized as shown in Fig.~\ref{Fig:hamiltonian_simulation_with_vtqg_simplified}.

\subsection{Pulse-efficient transpilation}

\begin{figure}[t]
    \includegraphics[width=8.0cm]{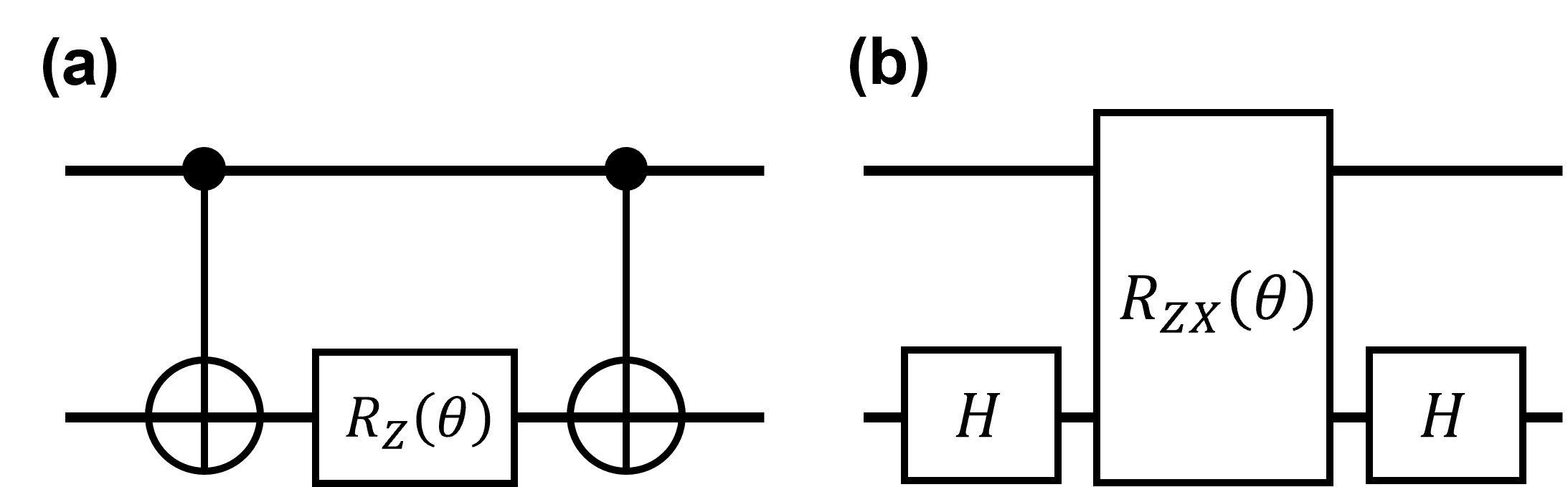}
    \caption{
    (a)A typical implementation of $R_{ZZ}(\theta)$ gate using two CNOT gates. (b) A pulse-efficient implementation of $R_{ZZ}(\theta)$ gate using a hardware-native $R_{ZX}(\theta)$ gate. 
    \label{Fig:hardware-native_gate_decomposition} 
    }
\end{figure}

In addition to the VTQG, in the experiment, we apply a pulse efficient transpilaiton to the quantum circuit. A pulse-efficient transpilaiton method enables a reduction in the number of two-qubit gates and the duration of their execution time, resulting in the better performance \cite{PhysRevResearch.3.033171, PhysRevResearch.3.043088, ferris2022quantum, kim2021scalable, niu2022effects, melo2022pulse}. Normally, an $R_{ZZ}(\theta)$ gate is implemented on quantum hardware using the decomposition shown in Fig.\,\ref{Fig:hardware-native_gate_decomposition}(a), where two CNOT gates are employed. By applying the pulse-efficient transpilaiton, an $R_{ZZ}(\theta)$ gate can be decomposed into a hardware-native $R_{ZX}(\theta)$ gate as shown in Fig.\,\ref{Fig:hardware-native_gate_decomposition}(b).

\section{Experimental results}\label{Sec:exp_res}

In the experiment, we set parameters $h=0.786$, $J=0.787$, and $dt=0.5$, respectively. We perform the experiments with a single Trotter step. In each experiment, we prepare three quantum circuits: (1) an original quantum circuit without any error suppression techniques, (2) a quantum circuit with a VTQG applied to the first qubit ($Q_0$) and the $N$th qubit ($Q_{N-1}$), (3) a quantum circuit with a VTQG and pulse-efficient transpilation.

\begin{figure*}[t]
    \includegraphics[width=17.0cm]{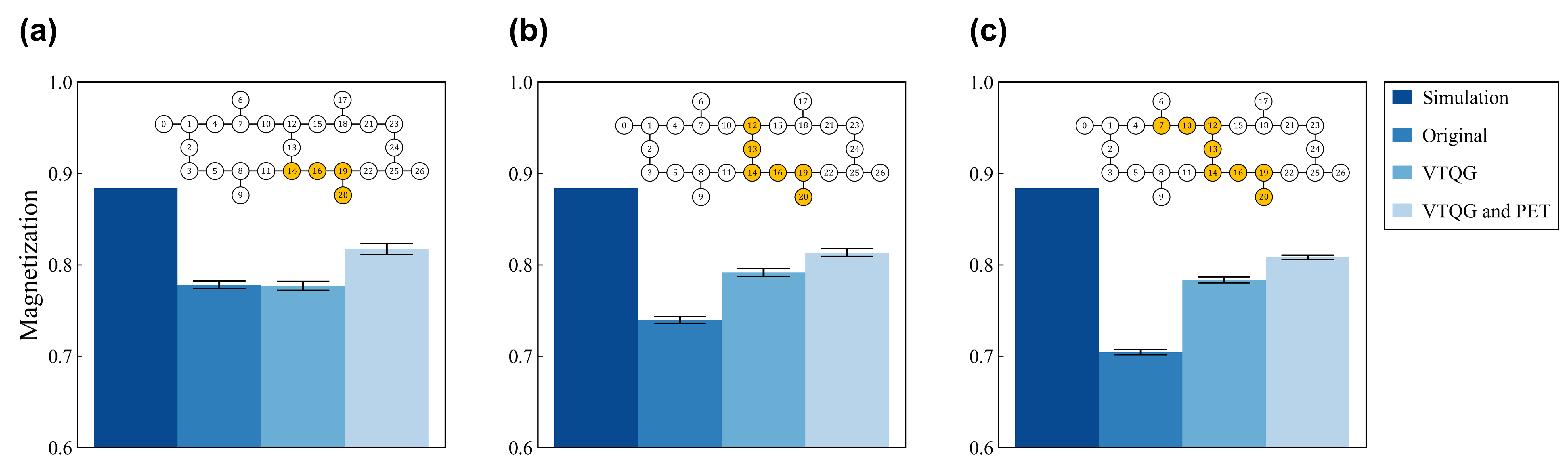}
    \caption{
    Dynamics of the one-dimensional transverse-field Ising model simulated with (a) 4, (b) 6, and (c) 8 qubits. The $y$-axis represents the magnetization, which is defined as Eq.\,(\ref{eq_mag}). The ideal value is calculated by using the statevector simulator in Qiskit. The result denoted as Original corresponds to the result obtained with a circuit without any error suppression techniques. The results obtained using either the VTQG or both the VTQG and the pulse-efficient transpilation (PET) are also shown. The inset illustrates the corresponding qubit layout used in each experiment. 
    \label{Fig:res_p1} 
    }
\end{figure*}

In Fig.\,\ref{Fig:res_p1}, the obtained results using a single Trotter step are given for $N=4, 6$ and $8$ qubits. The $y$-axis represents the magnetization, which is defined as 
\begin{equation}\label{eq_mag}
    \sqrt{\ev{\sigma_x}^2+\ev{\sigma_y}^2+\ev{\sigma_z}^2}.
\end{equation}
Here, each component is given as $\ev{\sigma_j}=\sum_{i=0}^{n-1}\ev{\sigma_j}_i/n$, where $j=x,y,z$. We also show the noiseless result, which is calculated using the statevector simulator in Qiskit \cite{Qiskit}.

We perform each experiment with 8192 shots. We also repeat each experiment 20 times and average over the 20 obtained results. The error bars given in Fig.\,\ref{Fig:res_p1} correspond to the standard deviation over these 20 instances.

We do not observe an improvement between the results with and without the VTQG for 4-qubit experiments while the performance improvement is observed for 6- and 8-qubit experiments. The difference can be explained by the number of SWAP gates required to apply a two-qubit gate between $Q_0$ and $Q_{N-1}$. We observe further improvement of the results between the experiments performed with quantum circuits with and without a pulse transplation method. This enhancement can be attributed to the reduction of the number of two-qubit gate operations and the overall gate execution time.

\section{Discussions}

One challenge of the application of the VTQG is the large overheads for its implementation. As the number of the VTQGs increases, the overheads for implementing the VTQG grows exponentially. Each VTQG requires execution of 10 circuits, hence, the number of circuits is $10^m$ for $m$ VTQGs. In fact, 100 quantum circuits are required to perform the experiments with two Trotter steps described in section\,\ref{Sec:exp_res}.

\begin{figure}[t]
    \includegraphics[width=7.0cm]{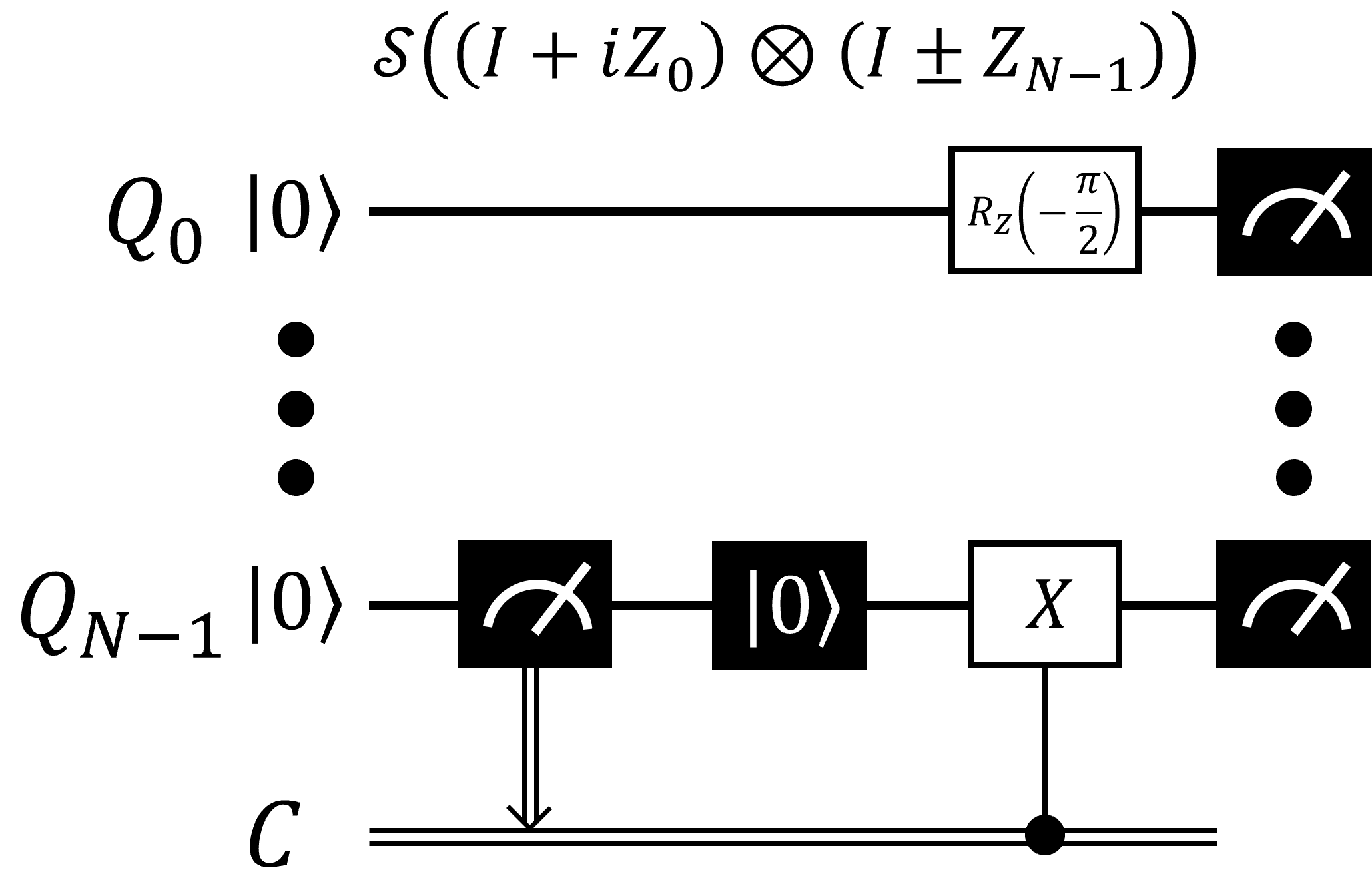}
    \caption{
    Projective measurement implemented using dynamic circuits. The result of a classical bit that registers the results of the projective-measurement on the qubit is used for deciding the sign of quasi-probability when classically post-processing the output distributions.
    \label{Fig:dynamic-circuit} 
    }
\end{figure}

A recent study, however, has shown that mid-circuit measurement enables an efficient implementation \cite{piveteau2022circuit}. In Fig.~\ref{Fig:dynamic-circuit}, we show an implementation of $S((I + i Z_0) \otimes (I \pm Z_{N-1}))$ terms of a VTQG for an $R_{ZZ}(\theta)$ gate using mid-circuit measurements, reset operations, and classical feedback. This implementation reduces the growth of the circuit to be executed from $\mathcal{O}(10^m)$ to $\mathcal{O}(6^m)$. Although we have not performed the experimental implementation using dynamic circuits in this study, mid-circuit measurement and the conditional operations based on the output of mid-circuit measurement are commonly used techniques \cite{PhysRevA.104.062440, govia2022randomized}. Therefore, this implementation scheme is feasible with the current level of quantum technologies.

Further reduction of errors may be achieved by adding additional different error suppression techniques. For instance, dynamical decoupling is used to achieve better qubit coherence \cite{PhysRevLett.82.2417, ezzell2022dynamical}. Pauli twirling can also be useful to suppress the coherent errors on a quantum device \cite{knill2004fault, PhysRevA.94.052325}. Furthermore, a scalable Hamiltonian simulation can be expected by combining a series of error suppression techniques with an error mitigation method \cite{PhysRevLett.119.180509, PhysRevX.7.021050, endo2021hybrid, cai2022quantum}, which has been demonstrated in Ref\,\cite{kim2021scalable}.

\section{Conclusions}

We have employed a virtual two-qubit gate (VTQG) as an error suppression technique for digital quantum simulation of the transverse-field Ising model to suppress the errors. The obtained results demonstrate that the VTQG can be used as a solid technique to suppress errors due to additional SWAP gates for the communication between distant qubits. Additionally, a further enhancement of the performance is observed by combining the VTQG with a pulse-efficient transpilation method.

%%%%%%%%%%%%%%%%%%%%%%%%%%%%%%%%%%%%%%%%%%%%%%%%%%%%%%%%%%%%
\begin{acknowledgments}
The authors thank Brian Quanz for a careful reading of the manuscript. For implementing the pulse-efficient transpilation, we used the functions given in qiskit-research repository \footnote{Qiskit Research, \url{https://github.com/qiskit-research/qiskit-research}, Accessed: 2022-12-16}.
\end{acknowledgments}

%%%%%%%%%%%%%%%%%%%%%%%%%%%%%%%%%%%%%%%%%%%%%%%%%%%%%%%%%%%%

\bibliography{main}

\end{document}